\documentstyle[aps,floats,epsf,twocolumn]{revtex}
\begin{document}
\newcommand{\be}{\begin{equation}}
\newcommand{\ee}{\end{equation}}
\newcommand{\bea}{\begin{eqnarray}}
\newcommand{\eea}{\end{eqnarray}}
\newcommand{\nn}{\nonumber}

\twocolumn[
\hsize\textwidth\columnwidth\hsize\csname@twocolumnfalse\endcsname
\draft

\title{Persistent currents in mesoscopic rings and boundary conformal field 
theory}
\author{S.\ Jaimungal, M.H.S.\ Amin and G.\ Rose}
\address{Department of Physics and Astronomy, University of British Columbia, 
Vancouver, BC V6T 1Z1, Canada}
\date{\today}
\maketitle

\begin{abstract}
A tight-binding model of electron dynamics in mesoscopic normal rings
is studied using boundary conformal field theory. The partition
function is calculated in the low energy limit and the persistent
current generated as a function of an external magnetic flux threading
the ring is found.  We study the cases where there are defects and
electron-electron interactions separately. The same temperature
scaling for the persistent current is found in each case, and the
functional form can be fitted, with a high degree of accuracy, to
experimental data.
\end{abstract}
\vspace{.5in}
]

\section{INTRODUCTION}

It is well known that the topology of a condensed matter system can
strongly influence its excitation spectrum. Consider for example the
textbook case of a superconducting ring. If we write the order
parameter in the form $\psi(x) \sim e^{i \Phi(x)}$ one can show that
there will exist a set of topological excitations characterised by the
winding number
\be
w={{1}\over{2\pi}} \oint_{\Gamma} \vec{\nabla} \Phi \cdot d\vec{l}
\ee
where $\Gamma$ represents a path encircling the hole in the
ring. These excitations in general carry a current due to the non-zero
gradient of the phase field in the ring. This effect is very
well-studied \cite{sc1}, being fundamentally responsible for much of
the physics of SQUIDs and related devices.

The prediction of the existence of this ``persistent current'' is,
however, not contingent upon the substrate material being
superconducting. Indeed it is apparent that the effect should be
present in any system with the topology of a ring that permits
excitations describable by field operators with the $U(1)$ symmetry
alluded to above. Persistent currents in normal systems have in fact
been observed--both in arrays of metal rings \cite{exp1} and in
individual metal \cite{exp3} and semiconducting \cite{exp5} rings.
From a theoretical point of view this is not surprising because the
persistent current is at heart an Aharonov-Bohm effect, independent of
the actual nature of the excitations in the material.

This being said, it is quite obvious that the {\it details} of the
supercurrent (such as its amplitude) will be strongly influenced by
the details of the sample under investigation.  For example, one
expects that the lifetime of the excitations carrying the current will
be of the order of the mean scattering time in the ring, which is
heavily dependent upon the density of scatterers in the material and
the details of the crystal field. If the associated mean free length
is much less than the length of the ring $L$, then the effect will
never be observed. Rigorous treatments of the effects of disorder on
the persistent current have been performed both numerically \cite{d1}
and analytically in restricted parameter regimes \cite{d3,d4,d5,d6}.

In addition to the effects of impurities, it is suspected that the
nature of the interaction between the supercurrent excitations can
drastically affect the details of the resultant current
\cite{i1,i2,i3}. For example, it has been shown that at zero
temperature and in one dimension, the amplitude to backscatter from an
impurity renormalizes to zero or to infinity for attractive or
repulsive inter-quasiparticle interaction respectively \cite{i4}.

In this paper we solve for the partition function of two different
models; one with arbitrary short-range electron-electron interaction
strength and no impurities, and one with an arbitrary number of
impurity sites but no interaction between electrons. These
calculations allow us to find the finite temperature persistent
current developed in both cases, which can be directly compared to
experiment \cite{exp3}. We find the temperature dependence of both
models, normalized to their zero temperature values for fixed external
flux, follows a universal form which agrees quantitatively with
experiment.

Our presentation is organised as follows. In section II we analyze the
case of a defect-free ring threaded by a magnetic flux with
short-range electron-electron Coulomb interactions. We demonstrate
that this problem can be solved via the use of boundary conformal
field theoretical techniques. In section III we consider a related
model, where the Coulomb interactions have been removed and an
arbitrary number of hopping defects added. It is shown that this new
model can be mapped to one of quasiparticles on a defect-free ring,
where the effective flux threading the ring is renormalized due to the
presence of the defects in the bare model.  In section IV we compare
our findings to zero-temperature theoretical results and the
finite-temperature data of Chandrasekhar et.al. \cite{exp3} with which
we find quantitative agreement. We conclude our presentation in
section V with a summary of our results.
 
\section{ANALYSIS OF DEFECT-FREE RING}

\subsection{Introduction of the Non-Interacting Model}

We begin our investigation by considering a tight-binding model in one
dimension with constant hopping integrals $t_0$ and no
electron-electron interaction. Throughout we shall choose a system of
units such that $\hbar=1$ and $k_B=1$. We allow the ring to be
threaded by a magnetic flux $\Phi=\int A_\phi dx = A_\phi L$, where
$A_\phi$ is the component of the vector potential along the ring. The
Hamiltonian for this model is
\be
H=  t_0 \sum_{j=1}^{N} \{ 
	e^{i \phi_0/N} \psi^{\dagger}_{j+1} \psi_{j}
	+ h.c. \}
\label{defH}
\ee
with a periodic boundary condition $\psi_{N+1}=\psi_{1}$.  Here we
define $\phi_0=2 \pi \Phi/\Phi_0$ with $\Phi_0=2 \pi c/e$ the standard
quantum of flux and $N$ the number of sites on the ring.  We assume
half-filling ($N$ even); extension to the case of $N$ odd is
straightforward.

\subsection{The Continuum Limit and Bosonization}

In this section we rewrite the model in a form that is accessible to
analysis by conformal field theoretical techniques. Because conformal
symmetry cannot exist in a lattice model, the first step is to rewrite
(\ref{defH}) as a continuum model.  We begin this procedure by first
``opening'' the ring and introducing an additional site at $j=N+1$. We
then perform a gauge transformation
\be
\psi_j \rightarrow e^{ i \phi_0 j /N} \psi_j
\label{gauge}
\ee
which allows (\ref{defH}) to be rewritten in the form
\be
H_0 = t_0 \sum_{j=1}^{N} \{ \psi^{\dagger}_{j+1} \psi_j +h.c. \}
\label{openH1}
\ee
In this case the periodic boundary conditions are changed to
\be
\psi_{N+1} = \kappa \psi_1
\label{bc0}
\ee
where $\kappa= e^{-i \phi_0}$.  We may now rewrite our model in terms
of left and right moving fermions, namely $\psi(x) = e^{i k_F x}
\psi_L(x) + e^{-i k_F x}\psi_R(x)$, where $x=j a$ in the lattice model. 
Taking the continuum limit in the standard way gives
\be
 H_0 = {{v_f}\over{2\pi}} \int_0^{L} dx
\left\{
\psi^{\dagger}_L(x)~i\frac{d}{dx} \psi_L(x) - \{L \leftrightarrow R\} \right\}
\label{h00}
\ee 
which is exact up to terms of the order of the lattice spacing $a$. For
ease of presentation we choose units such that 
the Fermi velocity $v_f=2 t_0 a =1$. It is now possible to bosonize (\ref{h00})
by introducing a field $\Phi$ such that
\bea
\psi_L(x) \sim e^{-i({{\Phi}\over{2r}}+2\pi r \widetilde{\Phi})}\quad, \qquad
\psi_R(x) \sim e^{i ({{\Phi}\over{2r}}-2\pi r \widetilde{\Phi})}
\label{psiLR}
\eea
on the interval $0 \le x \le L$. Here $\widetilde{\Phi}=\Phi_L-\Phi_R$
is the field dual to $\Phi =\Phi_L+\Phi_R$, and $r$ denotes the
compactification radius of the bosonic field $\Phi\sim\Phi+2\pi r$.
In the non-interacting case under consideration $r$ is taken to be at
its self-dual point $r=\frac{1}{\sqrt{4\pi}}$ . Note that it is not
necessary to include Klein factors \cite{klein} here as the number of
left and right moving fermions is conserved in this system.
Substitution of (\ref{psiLR}) into the boundary condition (\ref{bc0})
gives
\bea
\widetilde{\Phi}(L)-\widetilde{\Phi}(0) = {{\phi_0}\over{2\pi r}} \quad,\qquad
\Phi(L)+\Phi(0)=0 \label{bc1}
\eea
We introduce boundaries into the model by folding the ring in half.
Mathematically this is achieved by introducing two bosonic field
degrees of freedom defined on the half interval
\be
\Phi_{e,o}(x)= \Phi(L-x) \pm \Phi(x) \quad,\qquad 0 \le x \le L/2
\nn
\ee
In terms of the fields $\Phi_{e,o}(x)$ the boundary conditions
(\ref{bc1}) become
\bea
\widetilde{\Phi}_o(0) = {{\phi_0}\over{2\pi r}} \quad, \qquad
\Phi_e(0)=0
\label{Bosonbc}
\eea
and the Hamiltonian becomes
\bea
H_0 = {{1}\over{2\pi}} \int_0^{L/2} dx
~\left\{
\left({{d\Phi_e}\over{dx}}\right)^2 + 
\left({{d\Phi_o}\over{dx}}\right)^2 \right\}
\eea
We have reduced the boundary conditions on the fermions (\ref{bc0}) to
simple Dirichlet and Neumann conditions on the even and odd bosons
while maintaining the free nature of the Hamiltonian.

\subsection{The Role of Conformal Invariance}

Thus far we have translated our original problem of a perfect ring
threaded by a flux to one of two free bosons on the half-interval
$[0,L/2]$ with boundary conditions (\ref{Bosonbc}). The partition
function for such a system is immediately available via the use of
boundary conformal field theory techniques \cite{AfOs97}. In
this section we shall present an overview of the relevant conformal
machinery and compute the partition function for our model
(\ref{defH}).

In a conformally invariant theory, a conformal boundary condition is
given by those states which satisfy the condition $$
\left.\left (T - {\overline T} \right) \right|_{bd.} |B\rangle = 0 
$$ where $T$ and ${\overline T}$ denote the holomorphic and
anti-holomorphic components of the energy momentum tensor
respectively. This condition corresponds to the physical statement
that no energy is allowed to flow off of the edges of the system. The
general state satisfying the above condition is an Ishibashi state
\cite{Is89}, $$
|\mu \rangle\rangle \equiv \sum_{N} |\mu, N \rangle \otimes {\overline
{|\mu, N \rangle} } $$ where the sum is over all the descendents of
the heighest weight state defined by $|\mu\rangle$. In the case of a
single free boson, with compactification radius $r$, one can construct
states which not only satisfy the reflection condition but are also
eigenvectors of the field or its dual (with eigenvalue $\varphi_0$) at
the boundary \cite{AfOs97},
\bea
|D(\varphi_0)\rangle &=& \frac{1}{\sqrt{2r}}\sum_{k=-\infty}^\infty 
e^{-i k \varphi_0/r} \exp \left\{ - \sum_{n=1}^\infty a_n^\dagger 
{\tilde a}_n^\dagger \right\} ~|(0,k)\rangle \nn \\
|N(\varphi_0)\rangle &=& \sqrt{r}\sum_{w=-\infty}^\infty e^{-2irw \varphi_0}
\exp\left\{ +\sum_{n=1}^\infty a_n^\dagger 
{\tilde a}_n^\dagger \right\} ~|(w,0)\rangle \nn
\eea
These states correspond to Dirichlet and Neumann states
respectively. The details of the various operators appearing here can
be found in \cite{AfOs97}. It is sufficient to note that
$|(w,k)\rangle$ denotes a Fock vacuum with $a_0$ eigenvalue $k$ and
${\tilde a}_0$ eigenvalue $w$ and $a_n,{\tilde a}_n$ are the usual
annihilation operators which make up the bosonic field in canonical
quantization.

It is well known that one can compute the partition function of a
conformal field theory with boundary in two equivalent ways
\cite{Ca86,Ca89}: $Z=Tr e^{-\beta H_{ab}}$ where $H_{ab}$ denotes the
Hamiltonian which respects the boundary conditions $a$ and $b$ on
either end of the system, or through a modular transformation, which
interchanges space and time directions, as $Z=\langle a | e^{- L
H_\beta } | b \rangle$ where $H_\beta$ denotes the Hamiltonian on a
circle of radius $\beta$.  With this in mind, the single free boson
has four possible partition functions corresponding to the four
possible boundary conditions $DD, NN, DN$ or $ND$. These have all been
computed in \cite{AfOs97} and are given by
\bea
Z_{DD} &=& <D(\varphi_0)| e^{-{{1}\over{2}} H_{\beta}}|D(\varphi^{'}_0)> \nn \\
       &=& \frac{1}{\eta(q)}q^{-(\Delta\varphi_0/\pi)^2} 
\vartheta_3\left(-2 i r \Delta \varphi_0 \beta; q^{2r^2} \right) \nn \\
Z_{NN} &=& <N(\varphi_0)| e^{-{{1}\over{2}} H_{\beta}}|N(\varphi^{'}_0)> \nn \\
       &=& \frac{1}{\eta(q)}q^{-(\Delta\varphi_0/\pi)^2} 
\vartheta_3\left(-i \Delta \varphi_0 \beta/r; q^{1/2 r^2} \right) \nn \\
Z_{DN} &=& <D(\varphi_0)| e^{-{{1}\over{2}} H_{\beta}}|N(\varphi^{'}_0)> = 
\frac{1}{2\eta(q)}\vartheta_2(0;q) \nn
\eea
where $\Delta\varphi_0 = \varphi_0-\varphi_0'$ and $q=e^{-2\pi\beta}$
(The length scale $L$ has been absorbed in the definition of $\beta$,
i.e. to restore the length scales $\beta\to\beta/L$).  $\eta(q)$ and
$\vartheta_{2,3}(w,q)$ denote the Dedekind eta function and Jacobi
theta functions respectively. Explicitly these are
\bea
\eta(q) &=& q^{1/24} \prod_{n=1}^\infty (1-q^n) \nn\\
\vartheta_2(w,q) &=& 2 q^{1/4}\sum_{n=0}^\infty q^{n(n+1)}
\cos (2n+1)\omega \nn\\
\vartheta_3(w,q) &=& 1+2\sum_{n=1}^\infty q^{n^2} \cos 2n\omega \nn
\eea

\subsection{Calculation of the Partition Function}

It is now a straightforward matter to apply these results to the free
Hamiltonian for even and odd bosons with boundary conditions
(\ref{Bosonbc}).  Since the two bosons are non-interacting the
boundary states associated with them are tensor products of the single
boson boundary state. Therefore we can write the relevant partition
function as
\bea Z &=& \langle D(0) |\otimes \langle N(\phi_0/2\pi r)|
e^{-\frac{1}{2} (H_{\beta}^{e}+H_{\beta}^{o})} |N(0)
\rangle \otimes |D(0)\rangle \nn\\
&=& \langle D(0) |e^{-\frac{1}{2} H_{\beta}^{e}} |D(0)\rangle~ \langle
N(\phi_0/2\pi r)| e^{-\frac{1}{2} H_{\beta}^{o}} |N(0) \rangle \nn\\
&=& \frac{q^{-(\phi_0/2\pi^2 r)^2}}{\eta^2(q)}
\vartheta_3 \left(0;q^{r^2/2}\right)
\vartheta_3 \left(-{{i \phi_0 \beta}\over{\pi r^2}};q^{1/8 r^2}
\right)
\label{partn}
\eea
Notice that we have chosen the boundary state at $x=L/2$ to be
${\widetilde \Phi}_o = 0$ and $\Phi_e = 0$. One can be convinced that
this is correct by starting with the condition (\ref{bc0}) where the
sites $N+1$ and $1$ are replaced by $N/2+1$ and $N/2$ respectively and
taking $\kappa=1$. This effectively sets the flux in (\ref{Bosonbc})
to zero.

\subsection{Inclusion of Coulomb Interactions}

This completes the analysis of our original toy model
(\ref{defH}). Let us now consider what happens when we include
short-range Coulomb interactions. We repeat the procedure followed in
the previous sections with the Hamiltonian
\bea
H= &-& t_0 \sum_{j=1}^{N} \{ 
	e^{i \phi_0/N} \psi^{\dagger}_{j+1} \psi_{j}
	+ h.c. \} \nn \\
   &+& U_0 \sum_j \psi^{\dagger}(j)\psi(j)\psi^{\dagger}(j+1)\psi(j+1)
\label{defHU}
\eea
It is well known \cite{Af2} that the only effect of the Coulomb term 
is to renormalize both the radius of compactification
\be
r={{1}\over{\sqrt{4\pi}}} \sqrt{1+{{U_0}\over{t_{0} \pi}}}
\ee
and the Fermi velocity
\be
v_f = 2 t_0 a \left( 1 + {{U_0}\over{2 \pi t_0}} \right)
\ee
In order to accomodate this we choose new units such that $v_f=1$.  It
is then straightforward to use (\ref{partn}) to calculate the free
energy of the system and then the persistent current as a function of
the compactification radius $r(U_0)$. Explicitly this is
\be
J=-{{e}\over{\beta}} {{\partial \ln Z}\over{\partial \phi_0}}
\label{persist}
\ee
We show in fig.1 the results of this calculation. Varying the
parameter $r$ does not qualitatively affect the shape of $J(\phi_0,
\beta)$.

\begin{figure}[h]
\begin{center}
 \epsfxsize=75mm
 $$\epsffile{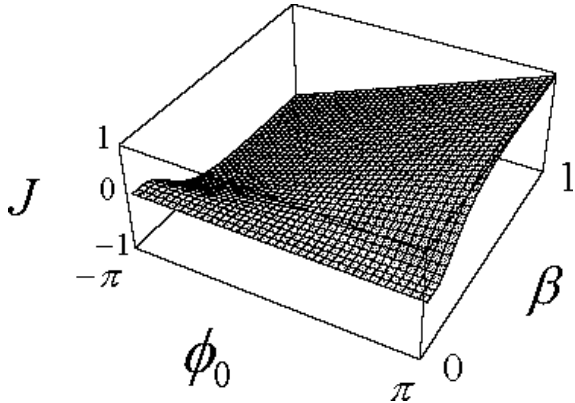}$$
\end{center}
\caption{Persistent current in a defect-free ring for arbitrary Coulomb 
interaction strength (units for $\beta$ are chosen to highlight the
transition from the high temperature regime to the low temperature
limit (corresponding here to $\beta \sim 1$) and units for $J$ are
arbitrary).  Note that $J$ is periodic in $\phi_0$ with period
$2\pi$.}
\label{fig1}
\end{figure}

\section{INCLUSION OF HOPPING DEFECTS}

We now demonstrate how this analysis may be extended to deal with the
presence of hopping defects.  The Hamiltonian for a ring with $m$
defects reads, after an appropriate gauge transformation,
\bea
H =&-& t_0\sum_{j\in\!\!\!| A} \left\{\psi^\dagger(j+1) \psi(j) + h.c.\right\} \nn \\
   &-& t_0\sum_{j\in A} \left\{\kappa_j~\psi^\dagger(j)\psi(j+1)+h.c.\right\} 
\label{defectH}
\eea
Here the set of sites $\{A\}$ labels the various defect links and
$\kappa_i = \rho_i\, e^{-i\phi_0/m}$ denotes the strengths of the
defects. It is a straightforward matter to diagonalize
(\ref{defectH});
\be
H_0= -\sum_{k:f(k)=0} \epsilon_k ~\eta^{\dagger}_k \eta_k
\label{QuasiH}
\ee
The dispersion relation here is $\epsilon_k\equiv 2 t_0 \cos(k)$, and
the restriction to $f(k)=0$ implements the quantization condition. 
Note that in all that follows we will only be interested in energies
very close to the Fermi surface. We shall assume that if there are
any localized states, their energies are far from $E_f$ and their
contribution to the low-energy physics can be neglected. 

As an explicit example of the quantization condition, one finds for the
case of a single defect $f(k)=0$ reduces to\cite{Henkel}:
\bea
f_1(k) &=& \rho^2 \sin(k (N-1)) \nn \\
       &+& 2\rho\cos(\phi_0) \sin(k)-\sin(k(N+1))=0
\label{f1k}
\eea
In the low-energy limit under investigation, for an arbitrary number of 
defects, the quantization condition
may be linearized near the Fermi points $k=\pm {\pi\over 2}$. Performing
this procedure, one finds that the effects of the defects and flux
may be subsumed into a shift in the fermionic spectrum given by
\be
k = \pm \frac{\pi}{2} + \frac{2\pi}{N} n + 
\frac{\alpha(\{ \kappa \},\phi_0)}{N} \quad , \qquad 
n\in {\rm Z\!\!Z} .
\ee
For the single defect case (\ref{f1k}) one can show \cite{Henkel} 
that $\alpha$ is given explicitly by
\be
\alpha= \cos^{-1} \left( {{2\rho}\over{1+\rho^2}} \cos \phi_0 \right)
\label{alpha}
\ee
In general for $m$ defects $\alpha$ is found by solving a secular
determinant of order $2^{m}$.

Introducing new quasi-particle fermionic operators
\be
{\widetilde\psi}(j) = \sum_{n}e^{i \frac{2\pi}{N} nj}~\eta_n
\label{tpsi}
\ee
into our Hamiltonian (\ref{QuasiH}) we find that
\be
H = -t_0 \sum_{j=1}^N\left\{ e^{i\alpha/N}
{\widetilde \psi}^\dagger(j+1){\widetilde \psi}(j) + h.c. \right\}
\label{newH}
\ee
which is the original toy model (\ref{defH}) with an effective flux
given by $\alpha$ instead of $\phi_0$.  This demonstrates that the
low energy limit of the multi-defect case can be reduced to one of free quasi-particles on a
ring threaded by an effective flux $\alpha(\{\kappa\},\phi_0)$. At
this point one can apply the analysis of the previous section to
obtain the partition function and therefore the persistent
current. Shown in fig.2 is the persistent current in the simplest case
where we have only one defect. We find that the flux dependence of $J$
changes dramatically as $\rho$ departs from $1$, evolving from the
sawtooth shape seen in the no-defect case to a sinusoidal variation.

\begin{figure}[h]
\begin{center}
 \epsfxsize=75mm
 $$\epsffile{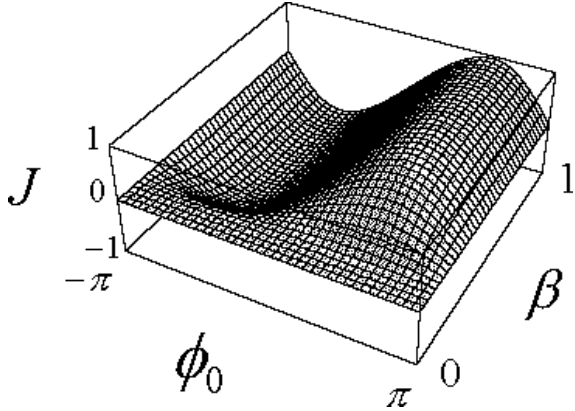}$$
\end{center}
\caption{Persistent current developed in a ring with no electron-electron
interaction and one hopping impurity with strength $\rho=6$. Units for $J$ and
$\beta$ are the same as those defined in figure 1.}
\label{fig2}
\end{figure}

\section{COMPARISON OF TEMPERATURE DEPENDENCE OF PERSISTENT CURRENT TO EXPERIMENT}

We have calculated the partition functions for both models under
consideration.  This allows the temperature dependence of the
persistent current to be extracted for each. We may perform both
calculations in parallel by retaining all factors of $r$ and $\alpha$
in our expressions, keeping in mind that these cannot be varied
independently. For the defect-free model, $\alpha = \phi_0$ and $r$ is
arbitrary.  For the non-interacting model, $\alpha$ is given by
solving for the roots of the quantization condition $f(k)=0$
(\ref{QuasiH}) linearized near the Fermi points and
$r=1/{\sqrt{4\pi}}$.
 
\begin{figure}[htbp]
\begin{center}
 \epsfxsize=75mm
 $$\epsffile{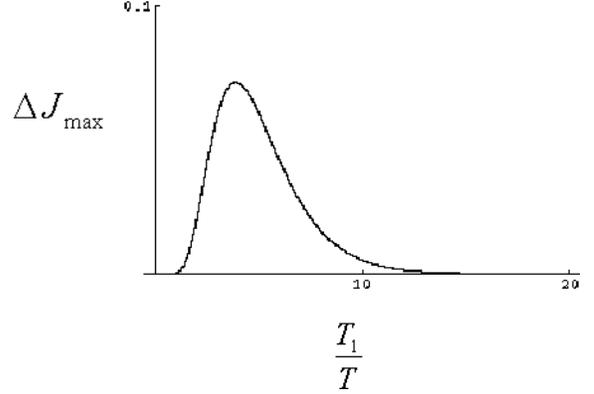}$$
\end{center}
\caption{$\Delta J_{max}$ is the maximum variation in $J/J_0$
for $\alpha \in [-\pi..\pi]$, here given as a function of $T_1/T$. 
We see that $\Delta J_{max} <<1$ for all $T_1/T$,
demonstrating the weak dependence of $J/J_0$ on $\alpha$.}
\vspace{.25in}
\label{fig3}
\end{figure}

With this caveat, the free energy can be written in general as
\bea
F &=& -{{1}\over{\beta}} \ln Z = {{\alpha^2}\over{2 \pi^3 r^2 }} - {{1}\over{\beta}} \left(
\ln \vartheta_3 (0; e^{-\pi \beta r^2}) \nn \right. \\
 &+& \left. \ln \vartheta_3 (-{{i \alpha \beta}\over{\pi r^2}}; e^{-4 \pi \beta/r^2})-
 2\ln\eta(q)  \right) \nn
\eea
with subsequent persistent current
\bea
J &=& -c {{\partial F}\over{\partial \vec{A}}} = -e 
{{\partial \alpha}\over {\partial \phi_0}} {{\partial F}\over{\partial \alpha}}
\label{J}
\eea
whose zero temperature limit is simply
\be
J_0= - {{e \alpha}\over{\pi^3 r^2}} {{\partial \alpha}\over{\partial \phi_0}}
\label{ZeroTJ}
\ee
If we normalize the persistent current to this value we find that
\be
{{J}\over{J_0}} = 1-{{i \pi^2}\over{\alpha}} {{\vartheta_3^{'}}\over{\vartheta_3}}
\left( {{i \alpha T_1}\over{2 \pi^2 T}},e^{-T_1/T} \right)
\label{iJ}
\ee
where we have introduced the energy scale $$ T_1={{4 v_f}\over{L r^2}}
$$ Note that though $\alpha$ can be a complicated function of $\phi_0$
and the defects, it is bounded, being defined in our model on the
region $[-\pi..\pi]$. It is interesting to note that the form of
(\ref{iJ}) is not a strong function of $\alpha$, its shape nearly
entirely dependent on the value of $T_1$ (see fig.\ref{fig3}). This means that regardless
of the model chosen, this one parameter (which can be varied
arbitrarily in both cases) determines the temperature dependence of
the normalized persistent current for fixed $\phi_0$, regardless of
the resultant value of $\alpha$.

\begin{figure}[htbp]
\begin{center}
 \epsfxsize=75mm
 $$\epsffile{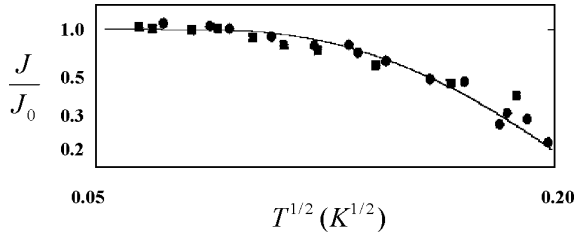}$$
\end{center}
\caption{Fit to the data of Chandrasekhar et.al.\protect\cite{exp3} with $T_1=0.09 K$.}
\label{fig4}
\end{figure}

We now compare the form derived for $J/J_0$ 
to the experimental results of Chandrasekhar
et.al.\cite{exp3} obtained from mesoscopic gold rings. In these
experiments the authors present data for the maximal normalized
persistent current, defined to be the solution of (\ref{iJ}) with
$\phi_0$ chosen so as to maximize the zero to peak amplitude of the
signal.  Shown in fig.(\ref{fig4}) are the results of a fit to the
data. The circles represent values gathered from a $1.4 \mu m$ x $2.6
\mu m$ loop and the squares the response from a $2.4 \mu m$ ring. We
find that the value $T_1=0.09 K$ fits both cases extremely well. This
finding lends credence to our claim that the details of the defects 
do not affect the temperature dependence of $J/J_0$. This is simply because
the two samples considered in the experiments must have had differing
defect structures (corresponding in our language to different shifts $\alpha$)
yet the experiments show the same temperature dependence.

\section{CONCLUSIONS}

Using boundary conformal field theoretical techniques we have computed the
partition function of a tight-binding model of electron dynamics in a
mesoscopic normal ring. Defects and
electron-electron interactions were introduced separately.  In the case where defects
are introduced, we mapped the problem onto a defect-free model with an
effective flux which depends on the details of the original defects.
The second case, where electron-electron interactions are included
within the defect-free model, was solved by noting that such
interactions serve only to alter the compactification radius of the bosons
and the effective Fermi velocity.

Using the partition functions we computed the persistent current for
both cases as a function of flux and temperature.  The model with
defects generated a persistent current that was found to change its flux
dependence from a sawtooth to a sinusoidal shape even for one weakly
scattering impurity. Futhermore, the amplitude of the oscillations
decreased as the strength of the defect increased. However, the
inclusion of electron-electron interactions in the defect-free case
produces a different behaviour: the sawtooth nature is not altered and
the amplitude of the current decreases only very weakly as the interaction strength is
increased. Although the functional dependence on flux differs in the
two cases described here, their normalized currents have {\it identical
temperature dependence}. There is one single free parameter in the
expression for the current which sets the scale in which temperature is measured. This
suggests that such a functional dependence on temperature is generic in the
sense that the microscopic details of the interactions and impurities
do not affect it when taken separately. 

We compared the temperature dependence with experiment\cite{exp3} and
with only one fitting parameter (the energy scale) found excellent
agreement. In such an experimental setup one expects that both
electron-electron interactions and defects play a significant role,
yet the predictions of our simple model were able to capture,
quantitatively, those results.  This seems to indicate that the
results obtained for the normalized persistent current
are not sensitive to any interplay between defects and electron-electron interactions
in the bare model--that is, including defects and electron-electron
interactions simultaneously should lead to analogous predictions.

\acknowledgements{We would like to thank I. Affleck, P. Stamp, A. Zagoskin,
J.S. Caux, C.L. Kane and Kurgan for helpful discussions and N. Prokofi'ev for
correspondence. This work was supported, in part, by the University of
British Columbia Graduate Fellowship program and by NSERC.}

\end{document}